\documentclass{article}

\usepackage{arxiv}

\usepackage[utf8]{inputenc} 
\usepackage[T1]{fontenc}    
\usepackage{url}            
\usepackage{booktabs}       
\usepackage{nicefrac}       
\usepackage{microtype}      
\usepackage{doi}

\usepackage{amsmath,amssymb,amsfonts}
\usepackage{algorithmic}
\usepackage{graphicx}
\usepackage{pgfplots}
\usepackage{pgf-pie}  
\usepackage{textcomp}
\usepackage{hyperref}
\usepackage{cleveref}
\usepackage[backend=biber, style=ieee]{biblatex}
\title{System and Method to Determine ME/CFS and Long COVID Disease Severity Using a Wearable Sensor}

\date{}

\author{
        \href{https://orcid.org/0000-0002-1591-7458}{\includegraphics[scale=0.06]{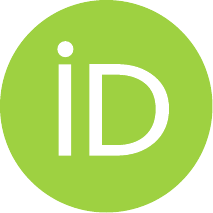}\hspace{1mm}Yifei Sun} \\
	Khoury College of Computer Sciences \\
	Northeastern University \\
	Boston, MA 02115 \\
	\texttt{ysun@ccs.neu.edu} \\
	\And
	\href{https://orcid.org/0000-0002-5241-8914}{\includegraphics[scale=0.06]{orcid.pdf}\hspace{1mm}Suzanne D. Vernon} \\
	Bateman Horne Center \\
	Salt Lake City, UT 84102 \\
	\texttt{sdvernon@batemanhornecenter.org} \\
 	\And
	\href{https://orcid.org/0000-0002-5256-628X}{\includegraphics[scale=0.06]{orcid.pdf}\hspace{1mm}Shad Roundy} \\
	Department of Mechanical Engineering \\
	University of Utah \\
	Salt Lake City, Utah 84112 \\
	\texttt{shad.roundy@utah.edu} \\
}

\renewcommand{\undertitle}{A Preprint}
\renewcommand{\headeright}{}
\renewcommand{\shorttitle}{System and Method to Determine ME/CFS and Long COVID Disease Severity Using a Wearable Sensor}

\begin{document}

\maketitle

\begin{abstract}
    Objective: We present a simple parameter, calculated from a single wearable sensor, that can be used to objectively measure disease severity in people with myalgic encephalomyelitis/chronic fatigue syndrome (ME/CFS) or Long COVID. We call this parameter UpTime. Methods: Prior research has shown that the amount of time a person spends upright, defined as lower legs vertical with feet on the floor, correlates strongly with ME/CFS disease severity. We use a single commercial inertial measurement unit (IMU) attached to the ankle to calculate the percentage of time each day that a person spends upright (i.e., UpTime) and number of Steps/Day. As Long COVID shares symptoms with ME/CFS, we also apply this method to determine Long COVID disease severity. We performed a trial with 55 subjects broken into three cohorts, healthy controls, ME/CFS, and Long COVID. Subjects wore the IMU on their ankle for a period of 7 days. UpTime and Steps/Day were calculated each day and results compared between cohorts. Results: UpTime effectively distinguishes between healthy controls and subjects diagnosed with ME/CFS ($\mathbf{p = 0.00004}$) and between healthy controls and subjects diagnosed with Long COVID ($\mathbf{p = 0.01185}$). Steps/Day did distinguish between controls and subjects with ME/CFS ($\mathbf{p = 0.01}$) but did not distinguish between controls and subjects with Long COVID ($\mathbf{p = 0.3}$). Conclusion: UpTime is an objective measure of ME/CFS and Long COVID severity. UpTime can be used as an objective outcome measure in clinical research and treatment trials. Significance: Objective assessment of ME/CFS and Long COVID disease severity using UpTime could spur development of treatments by enabling the effect of those treatments to be easily measured.
\end{abstract}


\section{Introduction}

Myalgic encephalomyelitis/chronic fatigue syndrome (ME/CFS) is a debilitating
disease with significant unmet medical needs that affects as many as 2.5
million people in the U.S. \cite{iom-mecfs-report} and causes enormous burden
for patients, their caregivers, the healthcare system and our society. The symptoms
of impaired physical function accompanied by severe fatigue, unrefreshing
sleep, cognitive impairment and orthostatic intolerance are worsened by
physical and cognitive exertion causing post-exertional malaise (PEM)
\cite{iom-mecfs-report}. As an under-studied disease which cause has yet to be
confirmed, it is estimated that between 84\% to 91\% of ME/CFS patients are not
yet diagnosed \cite{mecfs-factors}. At least one-quarter of ME/CFS patients are
house- or bed- bound at some point in their lives \cite{mecfs-pain-relief}. Due
to ME/CFS’s large patient population, affecting people working in various
fields, it costs the U.S. economy about \$17 to \$24 billion in medical bills and
lost income from lost household and labor force productivity per year
\cite{iom-mecfs-report}.

Biomarkers refer to a wide range of medical signs that can be objectively
measured and reliably reproduced. These medical signs
are distinguished from medical symptoms, which are restricted to the
indications of health or illness that patients perceive themselves
\cite{biomarker}. A subset of biomarkers, which can be measured using
digital tools, such as smartphones, wearable devices, and other digital sensors,
are called digital biomarkers. Digital biomarkers can be obtained in a
noninvasive and continuous manner, providing a wealth of data on
individuals' health status and disease progressions.

Long COVID is characterized by a group of symptoms that persist in some people following infection with SARS-CoV-2 \cite{Thaweethai2023-gs}. Studies report that the symptomotology between ME/CFS and Long COVID
overlaps significantly \cite{cdc-longcovid-about, symptomatology-mecfs-longcovid},
Thus, we hypothesize that biomarkers used for ME/CFS may show similar results in
diagnosing and measuring the severity of Long COVID.

Well-defined and reliable digital biomarkers are needed for objective diagnosis and to
measure clinically relevant and meaningful outcomes of treatment for ME/CFS and
Long COVID. Currently, diagnosis and treatment are challenging because there
are no specific biomarkers and tests for the disease. Widely used
diagnosis methods of the disease and similar symptoms in Long COVID patients are not based on statistically validated biomarkers, but on patients'
self-description of symptoms, questionnaires, and clinical observations
\cite{diagnostic-mecfs}. For example, in a recent CDC report, the suggested
diagnosis method is also largely based on patients' self reporting
\cite{cdc-mecfs-diagnosis}. The FDA published ``The Voice of the Patient
Report: Chronic Fatigue Syndrome and Myalgic Encephalomyelitis” meeting
transcript \cite{fda-voice-of-patient} noted that drug development for ME/CFS was not on the pharmaceutical industry
radar due to lack of valid, reliable, and measurable biomarkers and efficacy
endpoints that are critical for correct patient selection, drug dosing and
monitoring for safety and efficacy \cite{fda-workshop}, which underscores the need for development and regulatory acceptance of digital biomarkers for ME/CFS and similar diseases.

Based on clinical experience gathered from over 1,000 ME/CFS patients, studies at the Bateman Horne Center \footnote{The Bateman Horne Center is a non-profit
    research clinic specializing in the diagnosis and treatment of ME/CFS,
    fibromyalgia, post-viral syndromes, and related comorbidities based in Salt
    Lake City, UT, USA.} (BHC) indicated that patients' disease severity and degree of
physical impairment can be gauged by a survey based biomarker named
Hours of Upright Activity (HUA), which is defined as time spent with feet on
the floor, in an upright position, over a 24-hour period \cite{hua-oi-conenction}.
Severely ill ME/CFS patients reported 0 to 4 hours with their feet on the floor
while moderately ill patients reported having their feet on the floor for 5 to
8 hours. This observation led us to explore which ME/CFS symptoms were
associated with upright activity. Patients with less than 4 HUA over a 24-hour
period had significantly worse orthostatic intolerance symptoms ($p < 0.001$)
and significantly greater interference with walking and standing ($p < 0.001$)
compared to age and sex matched healthy controls \cite{sdv-oi}.
However, when symptoms begin to improve patients may return to premorbid
activity levels potentially causing PEM and disease relapse \cite{hua-uptime}.
These findings underscored the need for a precise, objective, and continuous measure 
of HUA to monitor disease severity and treatment efficacy. In response to this challenge,
this paper presents the development of Upright Position Time (UpTime) - a digital biomarker designed to passively 
and continuously record the duration for which a patient is in an upright position. The development of UpTime calculation and the results of an initial trial are reported in our previous study \cite{turner-mecfs}.

This work describes the further development of UpTime as a digital biomarker and the results of a clinical study to validate the measure as a digitial biomarker for ME/CFS so that it can be used to help monitor treatment response by physicians
and in clinical trials. Furthermore, given that UpTime is correlated with general activity levels, we evaluate if there is a correlation betweeen steps per day (Steps/Day) and UpTime. The remainder of this work is organized as follows: in the System and Methods section, we describe the hardware system (i.e., sensors, data collection, and storage) and software (i.e., parameter calculation and data processing pipeline) used to determine UpTime and Steps/Day. We then describe the clinical study undertaken in conjunction with the Bateman Horne Center to validate UpTime and Steps/Day as digital biomarkers. Finally, we present the results of the clinical study and discuss the prospects for both UpTime and Steps/Day as digital biomarkers for ME/CFS and Long COVID. 

\section{Systems and Methods}

\begin{figure}
\centering
\includegraphics[width=3.5in]{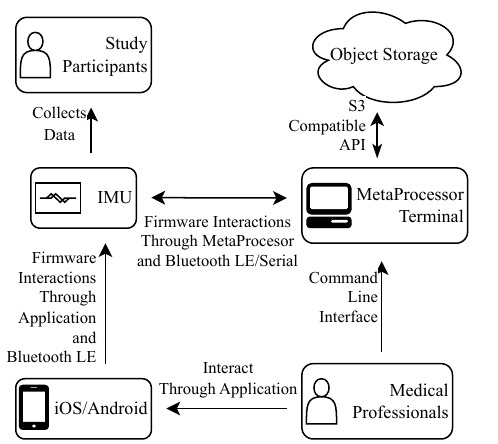}
\caption{System architecture: study participants (SP) goes to research clinic, clinicians set up MMS using the MetaBase app on an iOS or Android device, or MetaProcessor Terminal (CLI), then place MMS with ankle band onto SP. After SP mails MMS back to clinic, clinicians download data from MMS. The CLI compresses the data and sends it to object store, after which investigators can perform analyses on collected data.}
\label{fig:architecture}
\end{figure}

\subsection{Hardware System}


Both UpTime and and Steps/Day are calculated from raw data collected using 6-axis inertial measurement units (IMUs). In the previously mentioned pilot study \cite{turner-mecfs}, we used IMUs from Shimmer \cite{shimmer}. In that study, we required participants to wear a Shimmer IMU on the lateral side of each lower leg, approximately two inches above the malleolus (note that HUA and UpTime are defined as time spent upright, with feet on the floor, thus, only the orientation of lower limbs matters, not the orientation of the trunk. For further details on the definition and calculation of UpTime, we refer the reader to \cite{turner-mecfs, turner-thesis}). For the current study, we used MbientLab's MetaMotionS (MMS) \cite{mms} IMU sensor that is specifically designed for wearable applications, has more onboard data storage, and a longer battery life. Furthermore, we put only one IMU on the outer side of patients' lower leg as analysis of the data from the pilot study indicated that there was no significant difference in UpTime values calculated from a single leg as opposed to both legs \cite{turner-thesis}. The MMS features a 9-axis IMU that included accelerometer, gyroscope, and magnetometer. The MMS also features a barometer and temperature and ambient light sensors. However, we only made use of the accelerometer and gyroscope in this study. In logging mode, the MMS can last approximately 7 days. MbientLab provides comprehensive programming interfaces including low level C++ libraries to interact with the hardware and different bindings in other programming languages (Python, JavaScript, Swift) which provides significant customizability.

Previous research indicates that human motion is confined to low frequencies, specifically those below 10 Hz \cite{mms-sampling-human-motion}. To prevent aliasing, we chose to use a sample rate that is at least twice this limit \cite{mms-sampling-frequency} based on the Nyquist–Shannon sampling theorem. In practice, we selected a sample rate of 25 Hz in order to balance signal fidelity with battery lifetime and available onboard memory. The Bosch BMI160 chip embedded in the MMS, supports several different sensor ranges. Considering the nature of human motion and from experience gained from previous studies, the accelerometer was set to an output range of $\pm 8 g$, and the gyroscope was set to an output range of $\pm 1000 deg/sec$. These settings can be configured on the MMS device via Bluetooth from a smartphone (we used an iPhone 7 for this study), or via Bluetooth/Serial connection from a compatible computing device.

We use single-board computers, specifically the Raspberry Pi 3A and 4B units, to download data from the MMS IMUs, perform some minimal sensor fusion tasks, and upload the data to a cloud service provider. We use the Raspberry Pi as a cost effective method to utilize MbientLab's application programming interface (API) for Linux and to download data from multiple MMS sensors simultaneously and rapidly via wired connections.

\subsection{Software System: MetaProcessor}

We developed a custom data collection system with a modular architecture which we call MetaProcessor (See Figure \ref{fig:architecture}).
The preprocessor module takes in raw data from the IMU from one session, merges different sensor outputs and standardizes units. (One session includes 24/7 data collection over a 7 day period for one subject. Typically 1 session includes over 15 million rows of data where each row includes a time stamp, 3 accelerometer data points, and 3 gyroscope data points.) The preprocessing step includes interpolating any missing data (typically less than 10 data points per session) and time alignment as the IMU's internal timer is not synced with the Unix epoch. Clinical professionals manually record the start time, and the pre-processor aligns the dataset accordingly to ensure consistency and normalization for subsequent analysis.
The object store module automatically uploads preprocessed IMU data to a S3 compliant service provider and then deletes the local copy after checksum verification. This approach ensures that the data is safely stored, even in the event of device failures or data corruption. In this study, we have chosen Amazon Web Services Simple Storage Service (AWS S3) as our primary cloud object store service provider. The command line interface is invoked by the user and requires the user to supply relevant API configuration parameters, such as recording frequency, API endpoint, API keys, bucket name, and regular expressions for detecting incorrect file-naming (study participants ID, session ID, etc.). For more information about the configurations and usages of MetaProcessor, we refer readers to our GitHub repository \cite{metaprocessor}.

\subsection{Parameter Calculation}

The following MetaProcessor built-in modules calculate UpTime and Steps/Day from raw sensor data.

\begin{figure}
\centering
\includegraphics[width=3.5in]{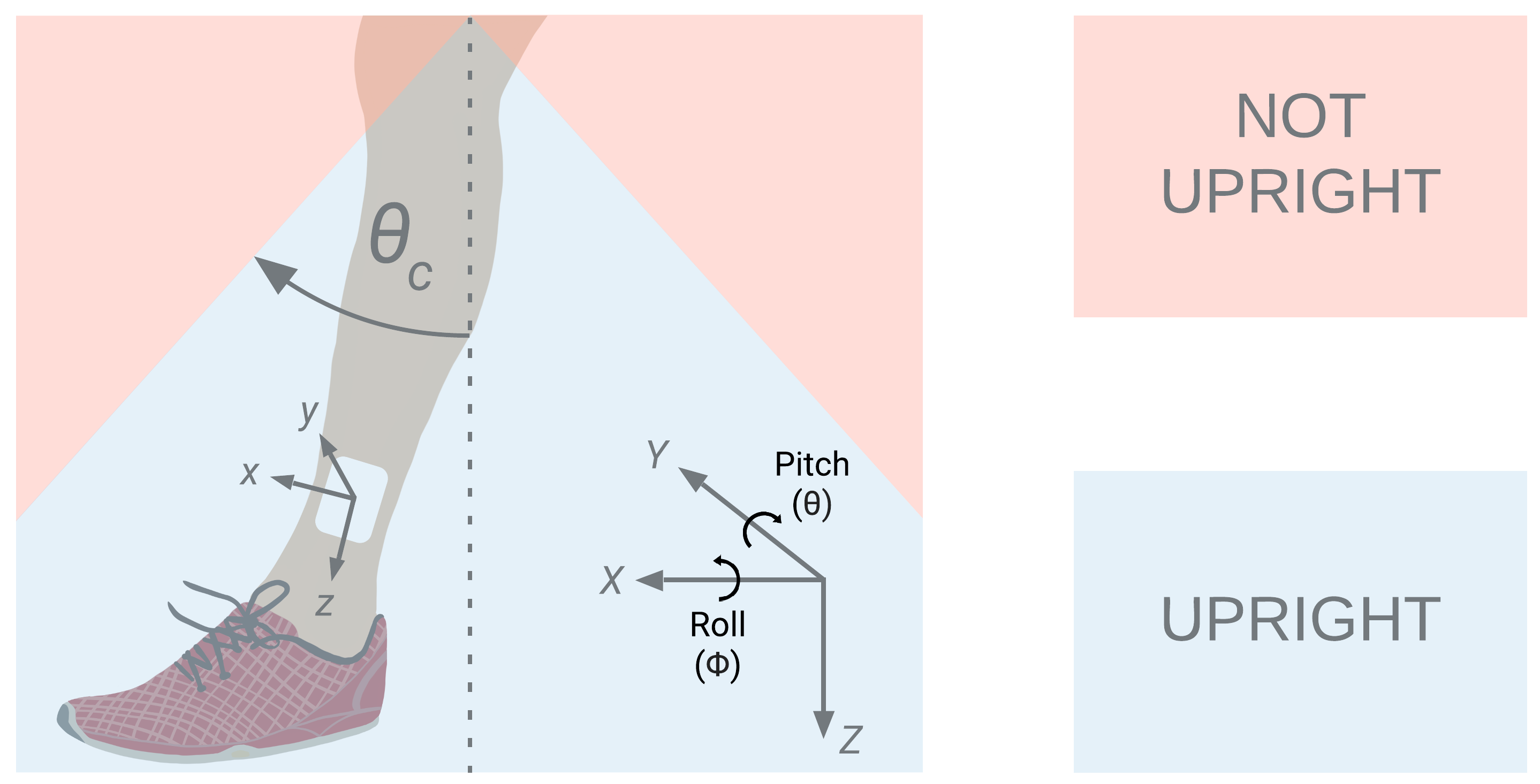}
\caption{The angle of each lower leg is compared to the critical angle $\theta_c$ to determine uprightness. A critical angle of 39$^{\circ}$ is used for this study. Reproduced from \cite{turner-mecfs}.  }
\label{fig:criticalAngle}
\end{figure}

UpTime: The UpTime algorithm, derived from \cite{turner-thesis}, calculates the amount of time a subject spends upright from the sensor data. As introduced in the Introduction, UpTime is an efficiently calculated parameter that correlates with patients' orthostatic intolerance level. Uptime is determined by first calculating the orientation of the lower leg for each sample (i.e. 25 times per second) and then determining if the lower leg is ``upright". To determine lower leg orientation, angle estimates derived from both the accelerometer readings and and gyroscope readings are optimally combined using a Kalman filter to minimize noise and bias error. The accuracy of lower leg orientations was verified in a pilot study in which subjects wore an IMU on each lower leg while performing a range of activities in a motion capture studio. If the angle of the lower leg (see Figure \ref{fig:criticalAngle}) is below a critical value, the subject is deemed to be "upright". A critical angle of 39$^{\circ}$ was determined based on clinical relevance and low sensitivity of Uptime to the critical angle. Further detail on the development and calculation of the Uptime parameter are available in \cite{turner-mecfs}, \cite{turner-thesis}.

Steps/Day: Another algorithm computes the step count for the subjects based on the accelerometer data. The algorithm uses the local variance method \cite{local-variance-1, local-variance-2, local-variance-3, local-variance-4, local-variance-5} to identify steps using the accelerometer data. The local variance method works by computing the local variance in the accelerometer data within a sliding window and detecting peaks that surpass a threshold value. These peaks correspond to the potential steps in the data. The algorithm can use either the x or y axis accelerometer. In practice, the algorithm outputs two step counts based on x- and y-axis for each accelerometer data set. The step count algorithm was validated using a preliminary study in which 5 subjects were asked to complete locomotion activities: a flat walk, a flat jog, and walking up and down stairs. Each subject was fitted with an IMU on each ankle, a Fitbit Inspire placed on each ankle right next to the IMU, and a third Fitbit Inspire device on the wrist. Steps were calculated from the raw x- and y-accelerometer data from each IMU and the three Fitbit devices returned step counts. These results were compared with the observed number of steps recorded by the experimenter. The step counts from the IMUs and the Fitbits worn on the ankle were similar resulting an average error of about 3\% for the flat walk, 7\% for the stairs, and 5\% for a flat jog. The Fitbit worn on the wrist resulted in significantly higher errors for all cases. It was deemed that as the performance of the step count algorithm was similar to a Fitbit worn on the ankle, it could be used as a reliable indicator of step count.   The Steps/Day parameter offers insights into the daily physical activity levels of the subjects, and is suspected to also correlate with the orthostatic intolerance level.

Additional time series feature extraction: MetaProcessor integrates a time series feature extraction library. The extracted features were not directly used in this study to evaluate ME/CFS or Long COVID disease severity. However, further studies utilizing extracted features could potentially result in even more robust digital biomarkers. 

Using MetaProcessor, we stably collected over 1 terabyte of raw IMU data securely with a very low error rate. In over 250 sessions only 3 errors occurred, 1 system error and 2 human errors, yielding less than 0.05\% system error rate. By providing an efficient, reliable, and user-friendly solution for managing IMU data, MetaProcessor has the potential to enhance the quality and efficiency of research projects in various fields, such as healthcare, human computer interaction, etc.

\subsection{Clinical Study}


UpTime was measured as part of a clinical investigation of endothelial function and upright activity in ME/CFS and Long COVID patients (WCG IRB \#20217163). We collected IMU data on 51 subjects of the 55 originally recruited . Of the 51 subjects, 30 were ME/CFS patients, 15 were Long COVID patients, and 6 were healthy volunteers.  

\begin{figure}
\centering
\includegraphics[width=3.5in]{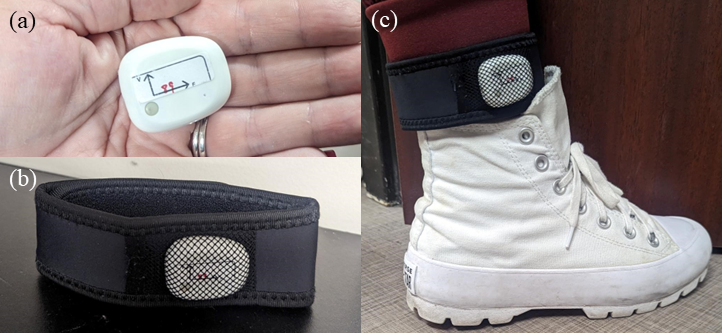}
\caption{Labeled IMU and deployment on study subject. (a) IMU device with axis direction markings. (b) IMU placed in flexible band to be worn on ankle. (c) IMU band on subject with IMU on the outside of the ankle.}
\label{fig:deviceOnSubject}
\end{figure}

Each subject in the study wore an MMS sensor on the outer side of their right ankle for 1 week continuously (see Figure \ref{fig:deviceOnSubject}). The only time the sensor was removed was during a bath or shower. The subjects were instructed to place the sensor in an orientation with the y-axis pointing upward if they were showering (so that the algorithm would record “upright”) or with the y-axis pointing horizontally if they were bathing (so that the algorithm would record “not upright”). The sensors were configured as described above in the Hardware System section, and data was processed utilizing a data collection station deployed at BHC. In addition, participants were asked to report the number of hours they spent upright each day (i.e., Hours of Upright Activity or HUA). 

At the end of the study, data from 3 subjects had to be discarded, leaving 48 subjects with valid data \footnote{Invalid entries: 1. Long COVID: END003 (setup error, accelerometer only), 2. ME/CFS: END006 (setup error, gyroscope only), 3. ME/CFS: END011 (bad firmware, the first downloading session with serial connection).}.

In addition to the data collected by the MMS devices, each patient underwent evaluations during their initial visit, including vital measurements. They also completed and submitted various survey forms online, and provided additional information about their health status. The vital measurements and survey data collected include: DANA Brain Vital (DANA), Augmentation Index (AI), Augmentation Index Normalized to HR 75 bpm (AI75), Baseline Heart Rate (bpm) (BLHR), Natural Base Log of Reactive Hyperemia Index (LnRHI), Reactive Hyperemia Index (RHI), and Hours of Upright Activity (HUA).

The integration of MMS sensor data with these additional health metrics provided a comprehensive dataset for the study, allowing researchers to analyze various aspects of the participants’ health and draw meaningful conclusions about the effectiveness of clinical trials. Although a broad dataset was collected, this work is more narrowly focused on the relationship between ME/CFS and Long COVID and the following three parameters: Uptime, HUA, and Steps/day. \footnote{The full dataset is too large for inclusion in this publication. It is available from authors upon request.}

\section{Results}

\begin{figure*}
\centering
\includegraphics[width=6.5in]{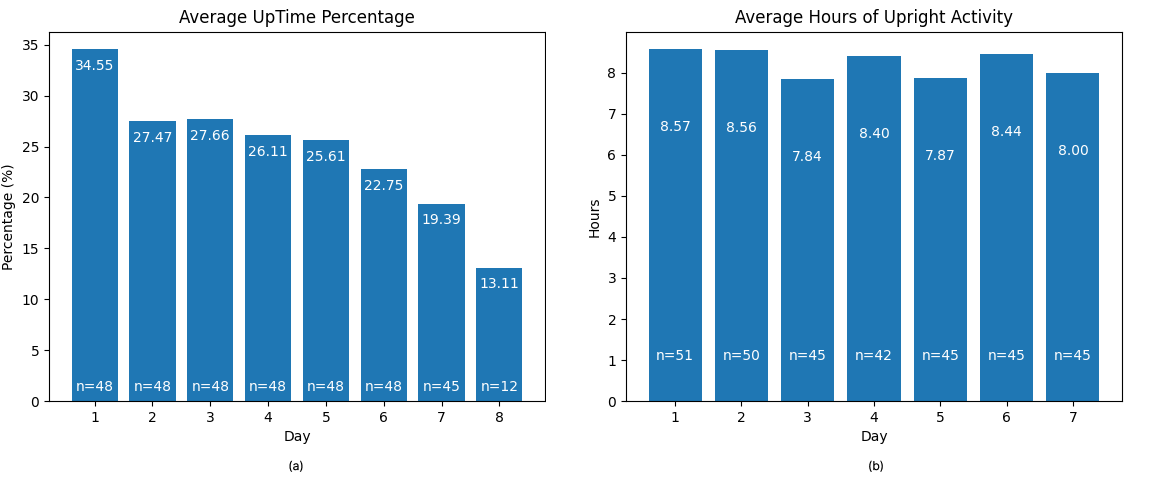} 
\caption{Average UpTime and HUA across cohorts. The variable $n$ represents the number of participants who contributed IMU data for a specific day, note that this does not imply a full 24-hour data collection for each participant. For example, in the UpTime graph, on day 6, $n = 48$, while on day 7, $n = 45$. This discrepancy indicates that 3 participants concluded their participation in the study on day 6, and as a result, we do not possess complete 24-hour recordings for these individuals on that day.}
\label{fig:average}
\end{figure*}

This section reports the results of several comparisons between different cohorts (ME/CFS, Long COVID, and health control) using UpTime, HUA, and Steps/Day.

Figure \ref{fig:average}(a) shows average UpTime each day for all cohorts. Average UpTime is reported as the percentage of time over a 24 hour period, starting at midnight, that the participant spends in an upright position as defined by HUA (see \cite{turner-mecfs} for a detailed description of how UpTime is calculated from raw IMU data). There is a substantially higher average on day 1 compared to subsequent days. We attribute this finding to the timing of the study, which typically commenced mid-day or in the afternoon. On day 1, participants arrived at the research clinic and were seated in an upright posture for an extended period, followed by traveling home, which also involved maintaining an upright posture (standing or sitting on public transport vehicle, or driving, or walking). Consequently a higher average UpTime is observed. Furthermore, days 7 and 8 exhibit a noticeably lower average UpTime. Participants were instructed to remove and return the IMU on day 7 via mail. Some subjects ($n = 12$) actually removed and returned the device on day 8. For the purposes of this study, a day begins at midnight. Therefore, on the final day of the study less than 24 hours of data are collected and the data that are collected are biased toward sleeping hours. Given these observations, we determined that days 2 through 6 provide the most consistent and relevant data for analysis. Therefore, the UpTime value for each study participant was calculated as the average from day 2 to day 6. This overall UpTime represents the participants’ overall uprightness score throughout the study’s duration. 

HUA is a self-reported measure that relies on patients logging into a secure portal and manually reporting their daily upright activity. This approach inherently introduces a degree of variability in the number of participants engaging in the survey throughout the study. Despite the consistent participation of all subjects on the first day, the number of participants fluctuates during the study’s course. This variation maybe attributed to several factors, including the convenience of the data reporting process, patients’ adherence to the protocol, or other external factors influencing individual engagement in the study. The daily average values derived from the HUA survey (see Figure \ref{fig:average}(b)) show a relatively smaller range of variation when compared with UpTime. The anomalies on days 1 and 7/8 are not present in the HUA results which would be expected as users are reporting their estimated hours of upright activity for the entire day. The daily average HUA ranges from 7.84 hours to 8.57 hours. 

We incorporated Steps/Day as a parameter in our study due to the ease of obtaining this data from readily available devices, such as smartphones and wearable technology. If Steps/Day is demonstrated to be a reliable digital-biomarker that can differentiate levels of orthostatic intolerance severity and subsequently distinguish ME/CFS severity, ongoing research efforts could be simplified by using it as an outcome measure. Similar to UpTime, days 1 and 7/8 show markedly different step counts than days 2 through 6 which are more consistent. Therefore, for similar reasons, Steps/Day is calculated as the average steps/day for each participant for days 2 through 6.

\begin{figure*}
\centering
\includegraphics[width=6.5in]{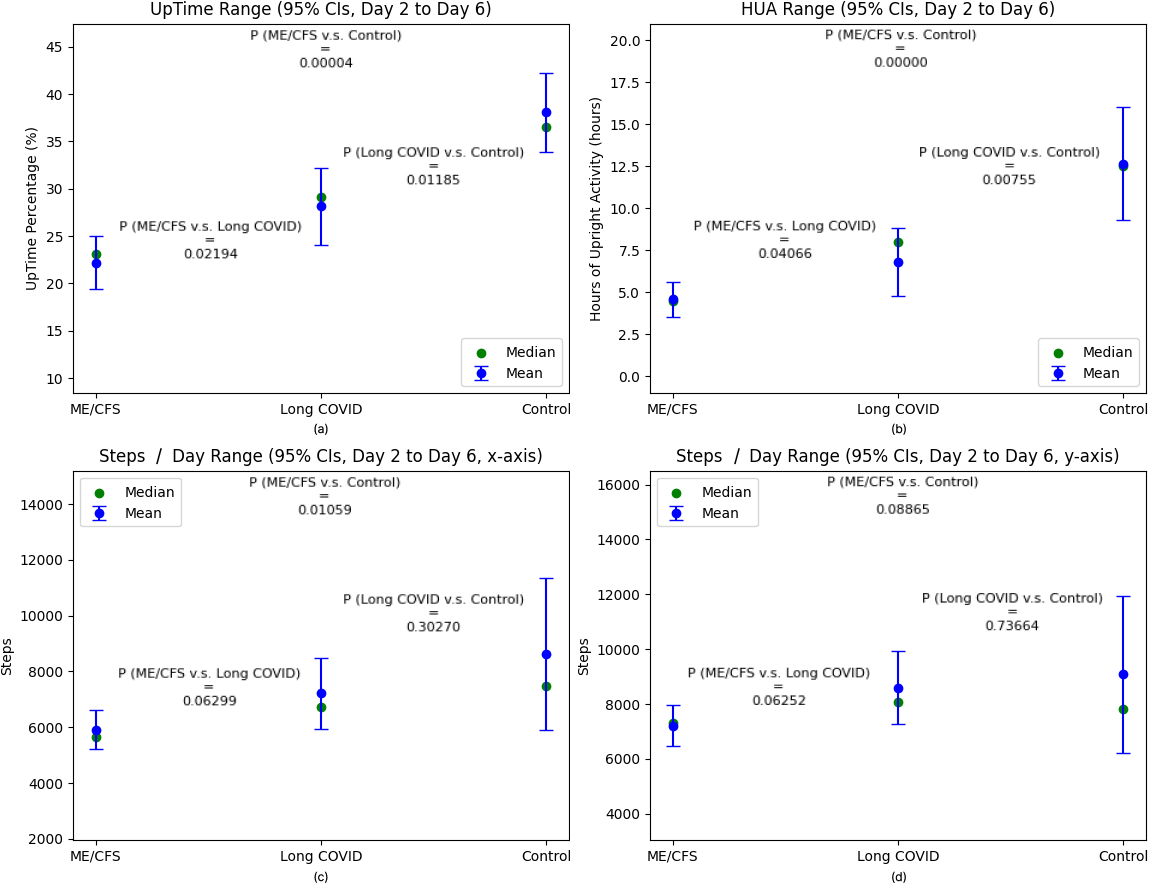}
\caption{The figure presents grouped t-tests performed at 95\% confidence level for UpTime (a), HUA (b) and Steps/Day (c, d) over a specified range of days . In the legend, the error bars represent the 95\% confidence intervals for UpTime, HUA, and Steps/Day of each group - ME/CFS, Long COVID, and Control. These intervals are calculated based on the standard error of the mean (SEM).} 
\label{fig:t-test}
\end{figure*}

We used multiple t-tests to assess the statistical significance of the differences in UpTime values between the three cohorts. Figure \ref{fig:t-test} shows the results of these multiple comparisons. 95\% confidence intervals for all three cohorts are shown for UpTime, HUA, and steps/day calculated with both x- and y-axes from the accelerometer. The p-values resulting from paired t-tests between cohorts are also shown on the graph.

In Figure \ref{fig:t-test}(a), the t-test comparing the UpTime of the ME/CFS cohort and the control cohort yielded a p-value of 0.00004, suggesting a strong difference between the two groups. Similarly, the t-test comparing the UpTime of the Long COVID cohort and the control cohort produced a p-value of 0.0119. Comparing the ME/CFS cohort to the Long COVID cohort, the t-test revealed a p-value of 0.0219. Applying the Bonferroni correction for multiple comparisons, a p-value below 0.0167 corresponds to a significance level of $\alpha = 0.05$. At this level, both ME/CFS and Long COVID groups differ significantly from the control group. The difference between the ME/CFS and Long COVID groups might be considered marginally significant. The 95\% confidence interval for the ME/CFS group is 19\% to 25\% UpTime, which corresponds well with the pilot study in \cite{turner-mecfs} which indicated a threshold of 30\% to distinguish between ME/CFS and control cohorts. The Long COVID cohort exhibited an UpTime range about midway between the ME/CFS and control cohorts.  

Similar to the UpTime results, the HUA t-tests demonstrated a significant difference between the ME/CFS vs. control groups ($p<0.000001$) and Long COVID vs. control groups ($p=0.007554$), but not between the ME/CFS and Long COVID groups ($p=0.0407$).  For the ME/CFS group, the variability in reported HUA is lower than UpTime (compare 95\% confidence intervals). However, for the control group, the variability in HUA is significantly larger than the variability in UpTime. Nevertheless, both measures appear to effectively differentiate healthy controls from participants with either ME/CFS or Long COVID. 

Broadly speaking, step counts did not significantly differentiate between cohorts. The step count data is quite similar regardless of whether the x- or y-axis accelerometer is used in the analysis. The one exception is that the p-value comparing ME/CFS vs. controls for the x-axis is 0.0106 while it is 0.0867 for the y-axis. This difference appears to be largely the result of the larger variance among healthy controls for the y-axis calculation. In fact, the large variation in steps/day for the healthy controls is what appears to limit the utility of this measure. Although the p-values could indicate some promise for steps/day as an outcome measure, both Uptime and HUA appear to be much better indicators.

We acknowledge that the step counts reported seem unrealistically large. The local variance algorithm depends on two related threshold parameters: the length of the sliding window from which local variance is calculated, and a peak threshold value. These two parameters were calibrated in a pilot study described above using healthy controls. The same threshold parameters were used in this study. As we did not perform a separate calibration and validation study with ME/CFS and Long/COVID patients, we cannot be certain that these threshold parameters are adequately calibrated. However, despite the large absolute values, we do still believe that the step counts do provide a useful relative measure of subject activity level.

\begin{figure}
\centering
\includegraphics[width=3.5in]{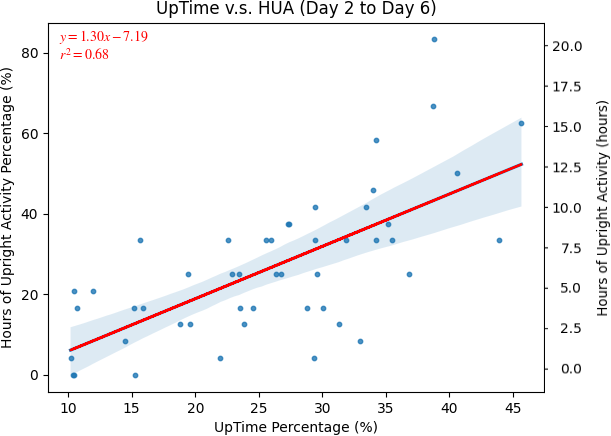}
\caption{Linear regression analysis between UpTime and Hours of Upright Activity.}
\label{fig:regression}
\end{figure}

Given the similarities between HUA and UpTime as measures of disease severity, we conducted a linear regression analysis to explore the relationship between these two parameters. As HUA is expressed in hours, while UpTime is represented as a percentage of total time in the day, we converted HUA to a percentage for the regression analysis.

Figure \ref{fig:regression} shows the results of the regression analysis. The two are moderately correlated ($r^2 = 0.68$) with a slope of 1.30. This moderately positive relationship may reflect the difference between self-reported experiences and objective measurements of disease severity. Interestingly, our analysis reveals that study participants are seemingly overestimating their HUA values in comparison to their corresponding UpTime data. However, at a closer look, there is a larger range of HUA values compared to UpTime (the HUA values range from 2\% to 60\% while the UpTime values range from 10\% to 45\%). This finding further highlights the potential discrepancies between subjective reporting and objective measurement. These insights emphasize the importance of utilizing both methods to gain a comprehensive understanding of patients’ disease severity and inform treatment approaches for ME/CFS and Long COVID patients.

\section{Discussion}

In the comparison between ME/CFS and healthy control groups, our findings can be summarized as follows. Both UpTime and HUA clearly distinguish participants with ME/CFS from healthy controls. This confirms prior findings from our pilot study, but with a larger dataset and lower p-values. Step counts did not distinguish between ME/CFS and control groups as clearly. Although the p-value of 0.0106 between groups using the x-axis accelerometer is significant, this was not replicated with the y-axis accelerometer and the p-value is much higher than for either UpTime or HUA. Initially we were hopeful that step counts could be reliably used as a measure of ME/CFS severity because it can be measured with such a wide range of commercially available wearables. However, in retrospect, it is not surprising that step count is a less effective outcome measure as it measures orthostatic intorlerance less directly than UpTime or HUA. Therefore, we do not recommend step count as a primary outcome measure for ME/CFS, although it could be incorporated with UpTime, HUA or other relevant parameters.

Our findings regarding the Long COVID group can be summarized as follows. As with ME/CFS, both UpTime and HUA significantly distinguished between participants with Long COVID and healthy controls. However, the significance levels were not as compelling (p-values of 0.012 and 0.0076) as for ME/CFS. The Long COVID group exhibited both UpTime percentages and HUA about midway between the ME/CFS and control groups. Both UpTime and HUA might be considered marginal differentiators between ME/CFS and Long COVID groups. However, this finding may not be very meaningful as the primary function of UpTime and HUA is to distinguished the severity of Long COVID or ME/CFS compared to healthy controls, not to distinguish between the two potential treatment groups. Finally, step count was not a reliable differentiator between Long COVID and control groups.

For this study, we logged raw IMU data at relatively high rates (i.e., 25 Hz) and we developed a full data pipeline (MetaProcessor) to manage and preprocess the data, and calculate relevant output parameters. The onboard data storage and battery life of the wearable (i.e., the MMS) are limiting factors on how long UpTime can be unobtrusively measured without manual data download and device recharging. However, we note that UpTime can be calculated efficiently in real-time on the wearable. And, in fact many IMU products now have built-in algorithms to determine orientation. Thus, in future iterations smaller devices and/or much longer maintenance-free monitoring are possible.

It is important to note that the root cause of Long COVID and ME/CFS is yet to be determined, and these two diseases might respond differently to the same treatment. Given the uncertainty surrounding the etiology of these conditions, we cannot make conclusive claims in the analysis of the Long COVID data. Our findings should be interpreted with caution, and further research is needed to better understand the underlying mechanisms and potential treatment strategies for both Long COVID and ME/CFS patients. Additionally, our study highlights the potential discrepancies between subjective reporting and objective measurements, emphasizing the importance of using both methods to gain a comprehensive understanding of patients’ disease severity. By examining these relationships and their implications, researchers can further refine their approaches to measuring and addressing disease severity in patients with ME/CFS and Long COVID, ultimately contributing to improved treatment outcomes and patient experiences.

\section{Conclusions}

The primary objective of this work was to design and demonstrate a wearable sensor-based system to collect raw IMU data over an extended period of time and compute higher level parameters/features (specifically UpTime and Steps/Day) with the ultimate goal of using the computed parameter to assess the severity of ME/CFS and Long COVID. A key element of the work was to develop a streamlined workflow that is scalable to large datasets involving many subjects with minimal intervention. Utilizing MbientLab's MetaMotionS IMU and our software system prototype, MetaProcessor, we collected and analyzed data from 48 subjects collected continuously over a 7 day study duration. The software pipeline achieved a system error rate of less 0.05\%. The data resulting from the study demonstrated that UpTime is an effective and objective digital biomarker for both ME/CFS and Long COVID disease severity. The study also demonstrated that step counts alone are not an effective digital biomarker for either condition.  Upon further validation, UpTime could be confidently proposed as an objective clinical measure for disease severity classification in the context of ME/CFS and related medical conditions. Additionally, HUA and Steps/Day could serve as supplementary biomarkers to further enhance the accuracy of the assessment. Future research could explore the integration of multiple digital biomarkers and potential confounding factors to improve the accuracy and reliability of these measures in the context of orthostatic intolerance severity among patients with ME/CFS and Long COVID.

\section*{Acknowledgement}

We would like to acknowledge the Solve ME/CFS Foundation (Grant No. 10054190) and Dr. Sonak Pastakia, Purdue University, for providing the funding to conduct this study. We gratefully acknowledge the Bateman Horne Center for supporting this study. We would especially like to thank Candace Rond and Sarah Schieving for all their help organizing the study and collecting data.

\renewcommand*{\UrlFont}{\rmfamily}
\printbibliography

\end{document}